\documentclass[prb,aps,showpacs,floatfix,floats,nobalancelastpage,twocolumn]{revtex4-1}
\usepackage{graphicx}
\usepackage{graphics}
\usepackage{latexsym,amsmath,amssymb,bm,euscript}
\usepackage{color}
\usepackage{dcolumn}
\usepackage{bm}
\usepackage{epstopdf}
\usepackage{times}
\usepackage{multirow}
\usepackage{wasysym}
\usepackage{subfigure}

\def\etal{~\textit{et~al.}}
\def\ra{\rangle}
\def\la{\langle}
\def\up{\uparrow}
\def\dn{\downarrow}
\def\Hc{{\rm H.c.}}

\def\ET{{$\kappa$-(ET)$_2$Cu$_2$(CN)$_3$}}
\def\dmit{{EtMe$_3$Sb[Pd(dmit)$_2$]$_2$}}

\begin{document}

\title{Power-Law Behavior of Bond Energy Correlators in a Kitaev-type Model with a Stable Parton Fermi Surface}
\author{Hsin-Hua Lai}
\affiliation{Department of Physics, California Institute of Technology, Pasadena, California 91125, USA}
\author{Olexei I. Motrunich}
\affiliation{Department of Physics, California Institute of Technology, Pasadena, California 91125, USA}
\date{\today}

\begin{abstract}
We study bond energy correlation functions in an exactly solvable quantum spin model of Kitaev type on the kagome lattice with stable Fermi surface of partons proposed recently by Chua \etal, Ref.\ [arXiv:1010.1035].  Even though any spin correlations are ultra-short ranged, we find that the bond energy correlations have power law behavior with a $1/|{\bm r}|^3$ envelope and oscillations at incommensurate wavevectors.  We determine the corresponding singular surfaces in momentum space, which provide a gauge-invariant characterization of this gapless spin liquid.
\end{abstract}
\maketitle

\section{Introduction}
In the last two decades, there has been dramatic theoretical progress in our understanding of RVB ideas\cite{PWA_science} and spin liquids.\cite{LeeNagaosaWen, Lee08_science, Balents_nature}  We now know that there are many different kinds of spin liquids.  Gapped topological spin liquids\cite{Kalmeyer87, KivRokSet87, Read89, ReadSachdev, Wen91, SenthilFisher_Z2, MoessnerSondhi, WenPSG, Kitaev06} are best understood and have been shown to exist in model systems.  Gapless spin liquids are also possible\cite{WenPSG, SenthilFisher_Z2, LeeNagaosaWen} and recently realized in experiments\cite{Shimizu03, Kurosaki05, SYamashita08, MYamashita09, McKenzie, Itou07, Itou08, Itou10, MYamashita10, Powell10}, but are understood to a lesser degree, particularly when both the emergent parton and gauge field excitations are gapless.\cite{LeeNagaosaWen, Polchinski94, Altshuler94, YBKim94, SSLee2009, Metlitski2010, Mross2010, DBL, Rantner2002, Hermele2004} 

From the early days, slave particle approaches\cite{LeeNagaosaWen} have played an important role in studying such phases.  The discovery by Kitaev\cite{Kitaev06} of an exactly solvable, interacting two-dimensional spin-1/2 model on the honeycomb lattice with a spin liquid phase paved an exciting road for the study of spin liquids.  Since then, there have been many studies of Kitaev-type models.\cite{Wen03, Feng07, Baskaran07, Lee07, Yao07, Chen07, Vidal08, Yao09, Mandal09, Wu09, Nussinov09, Baskaran09, Willans10, Tikhonov10, Tikhonov10-arxiv, Chern10, Wang10, Chua10,Yao10, Dhochak10}.  In the last year, several of them realized gapless spin liquid phases with {\it parton Fermi surfaces}\cite{Yao09, Baskaran09, Tikhonov10, Chua10} (and gapped $Z_2$ gauge fields).  In this work, we want to directly detect the presence of a surface of low-energy excitations.  Note that the parton Fermi surface itself is gauge-dependent and is not accessible via local observables.  However, there is a geometric surface information that is physical and can be detected using gauge-invariant local energy observables.  We take up a very recent model by Chua\etal\cite{Chua10} to illustrate this point.

Chua\etal\cite{Chua10} proposed an exactly solvable spin-3/2 model on the kagome lattice and found a regime with a gapless spin liquid with a {\it stable} Fermi surface.  Motivated by such a spin liquid phase and known techniques to characterize situations with gapless partons,\cite{LeeNagaosaWen, Altshuler94, DBL, Rantner2002, Hermele2004} we propose to study {\it gauge-invariant} operators such as bond energy operators.  Our main results show that, unlike any spin correlations which are ultra short-ranged, the bond energy correlations have {\it power-law} behaviors with $1/|{\bm r}|^3$ envelope in real space and also oscillations at incommensurate wavevectors which form what we call {\it singular surfaces}\cite{LeeNagaosaWen, Altshuler94, DBL} in the momentum space. An interesting aspect of the Chua\etal\cite{Chua10} model is that there is no ``nesting'' in the Cooper channel for the low-energy fermions because of the absence of inversion symmetry and the broken time-reversal.  This gives a non-trivial ${\bm k}_{FR} + {\bm k}_{FL}$ critical surface in addition to more familiar ${\bm k}_{FR} - {\bm k}_{FL}$ (a.k.a.\ ``$2k_F$'') surface in the local energy correlations.

In connection to experiments, the physics discussed here can be conceptually related to the recent gapless spin liquids in the organic compounds \ET  and \dmit,\cite{Itou07, Itou08, Itou10, MYamashita10} where much thinking focused around the possibility of gapless Fermi surface of spinons.  Although the Kitaev-type theoretical models are not directly appropriate for these materials, some of the qualitative physics discussed here applies more generally to gapless spin liquids and has practical implications.  Thus, such 2${\bm k}_{F}$ physics information can also be revealed by measuring Ruderman-Kittel-Kasuya-Yosida (RKKY) interaction between magnetic impurities\cite{Dhochak10, Norman2009} or by measuring textures in local spin susceptibility (Knight shift experiments) or other local properties near a non-magnetic impurity.\cite{Willans10, Lai2009}  We will discuss further possible connections in the conclusion section.

We also mention that entanglement properties of a ground state wavefunction can be used for characterizing a phase of matter, especially for gapless spin liquids, in addition to gauge-invariant observables with power law correlations.  For instance, a recent paper\cite{Zhang11} measured entanglement entropy in the Gutzwiller-projected Fermi sea wavefunction on the triangular lattice and found logarithmic violation of the area law, which strongly suggests the existence of gapless Fermi surface in the resulting spin liquid state.  Other recent works\cite{Block2010} used the entanglement entropy to estimate the central charge in DMRG studies of spin-1/2 Hamiltonians with ring exchanges on multi-leg ladders, and found that the central charge increases with the number of legs as expected in such gapless spin liquids.

The paper is organized as follows.  In Sec.~\ref{Sec:Kitaev-type H} we start from the Chua\etal\ Hamiltonian\cite{Chua10} on the kagome lattice.  In Sec.~\ref{Sec:Bond correlator} we define bond energy correlation function.  In Sec.~\ref{Sec:Long-wavelength analysis} we provide a theoretical approach to describe the long-distance behavior of the correlations.  In Sec.~\ref{Sec:Numerical data} we present exact numerical calculations of the bond energy correlations.  We conclude with some speculations about similarity with recent experiments in \dmit\ in which a gapless spin liquid has been realized.\cite{Itou07, Itou08, Itou10, MYamashita10}

\section{Chua-Yao-Fiete Kitaev-type Hamiltonian}\label{Sec:Kitaev-type H}
\begin{figure}[t]
\includegraphics[width=\columnwidth]{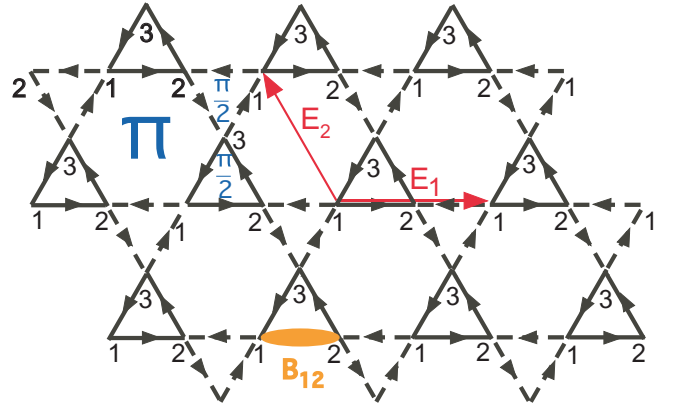}
\caption{
Kagome lattice with three sites (labelled as 1, 2, and 3) per unit cell.  We consider Chua\etal\cite{Chua10} model with the ground state flux configuration as shown, $\{ \phi_{\hexagon}, \phi_\triangle, \phi_\nabla \} = \{\pi, \pi/2, \pi/2\}$.  We fix the gauge by taking $i u_{ij} = i$ with bonds directed counter-clockwise around the triangles.  We also show the bond energy operator $\mathcal{B}_{12c}$ whose correlations are presented in this paper, while other bond energy operators have qualitatively similar correlations. 
}
\label{Kagome Lattice}
\end{figure}

We begin by formulating the Hamiltonian in the parton language.  The model is defined on the kagome lattice, see Fig.~\ref{Kagome Lattice}.  On each site $i$ of the kagome lattice, there is a physical four-dimensional Hilbert space realized using six Majorana fermions $\xi^1_i$, $\xi^2_i$, $\xi^3_i$, $\xi^4_i$, $c_i$, and $d_i$, with the constraint $D_i \equiv -i\xi^1_i \xi^2_i \xi^3_i \xi^4_i c_i d_i = 1$ (namely, for any physical state $|\Phi\ra_{\rm phys}$, we require $D_i |\Phi\ra_{\rm phys} = |\Phi\ra_{\rm phys}$).  The Chua-Yao-Fiete Kitaev-type Hamiltonian is
\begin{eqnarray}
\mathcal{H} &=& i \sum_{\la ij \ra} u_{ij}\left[ J_{ij} c_i c_j + J_{ij}^{\prime} d_i d_j \right] + i J_5 \sum_i c_i d_i \hspace{0.3cm}\label{kitaev H}\\
&& -\alpha \sum_{\hexagon} W_{\hexagon} - \beta\sum_{\la \triangle, \nabla \ra} W_\triangle W_\nabla ~, \label{stablize flux}
\end{eqnarray}
where $\la ij \ra$ represents nearest neighbor links and $u_{ij} = -u_{ji}$, with $u_{ij} \equiv -i \xi^1_i \xi^2_j$ if $\la ij \ra \in \triangle$ and $u_{ij} \equiv -i \xi^3_i \xi^4_j$ if $\la ij \ra \in \nabla$ for bond directions chosen to go counter-clockwise around the triangles.  Placket operators $W_p = \prod_{\la ij \ra \in p} u_{ij}$, with $p = \triangle,~\nabla,~\hexagon$, are gauge-invariant (i.e., act in the physical Hilbert space) and are conserved by the Hamiltonian.  The terms in Eq.~(\ref{stablize flux}) with $\alpha>0$  and $\beta>0$ are added to stabilize particular  
ground states with $W_{\hexagon} = 1$, $W_\triangle = W_\nabla = \pm 1$. Since in the Kitaev-type model, $[u_{ij}, \mathcal{H}] = [u_{ij}, u_{i'j'}] = 0$, we can treat the $\mathbb{Z}_2$ gauge fields $u_{ij}$ as static background and replace by their eigenvalues $\pm 1$.  We then have free Majorana fermions $c$ and $d$ hopping on the lattice in the presence of ``fluxes'' $\phi_p$ defined via $e^{-i\phi_p} \equiv \prod_{\la ij \ra \in p} i u_{ij}$.

Throughout, we work in the ground state with $W_{\hexagon} = 1$, $W_\triangle = W_\nabla = 1$ which breaks time reversal symmetry; this translates to fluxes $\{ \phi_{\hexagon}, \phi_\triangle, \phi_\nabla \} = \{\pi, \pi/2, \pi/2\}$ as shown in Fig.~\ref{Kagome Lattice}.  We fix the gauge by taking $u_{ij} = 1$ with bonds $i\rightarrow j$ directed counter-clockwise around the triangles.  There are three physical sites per unit cell and six remaining Majoranas per unit cell.  We replace the labeling $\{c_i, d_i\}$ with $\Psi^M_{I=\{{\bm r},a\}}$, where ${\bm r}$ runs over the Bravais lattice of unit cells of the kagome network and $a$ runs over the six Majoranas in each unit cell (three $c$ Majoranas and three $d$ Majoranas).  The Hamiltonian can be written in a concise form,
\begin{eqnarray}
\mathcal{H} &=& \sum_{\la ({\bm r},a), ({\bm r}',a') \ra} \Psi^M_{{\bm r},a} A_{{\bm r},a;~{\bm r}',a'} \Psi^M_{{\bm r}',a'} \\
&=& \sum_{\la IJ \ra} \Psi^M_I A_{IJ} \Psi^M_J ~. \label{Kitaev-type H}
\end{eqnarray}
The Majorana field satisfies the usual anticommutation relation, $\{ \Psi^M_{{\bm r},a}, \Psi^M_{{\bm r}',a'}\} = 2 \delta_{{\bm r}{\bm r}'} \delta_{aa'} = 2\delta_{IJ}$.  In the chosen gauge, there is translational symmetry between different unit cells; hence, $A_{{\bm r},a;~{\bm r}',a'} = A_{aa'}({\bm r} - {\bm r}')$.

In order to give a concise long wavelength description, it will be convenient to use familiar complex fermion fields.  To this end, we can proceed as follows.  For a general Majorana problem specified by matrix $A_{IJ}$, we diagonalize $A_{IJ}$ for spectra, but only half of the bands are needed while the rest of the bands can be obtained by a specific relation and are redundant.  Explicitly, for a system with $2m$ bands, we can divide them into two groups.  The first group contains bands from $1$ to $m$ with eigenvector-eigenenergy pairs $\{\vec{v}_{b,{\bm k}}, \epsilon_{b,{\bm k}}\}$, where $b = 1, 2, \dots, m$ are band indices, and the second group contains bands from $m+1$ to $2m$ related to the first group, $\{\vec{v}_{b'=m+b, {\bm k}}, \epsilon_{b'=m+b, {\bm k}}\} = \{ \vec{v}^*_{b, -{\bm k}}, -\epsilon_{b, -{\bm k}}\}$.

\begin{figure}[t]
\includegraphics[width=\columnwidth]{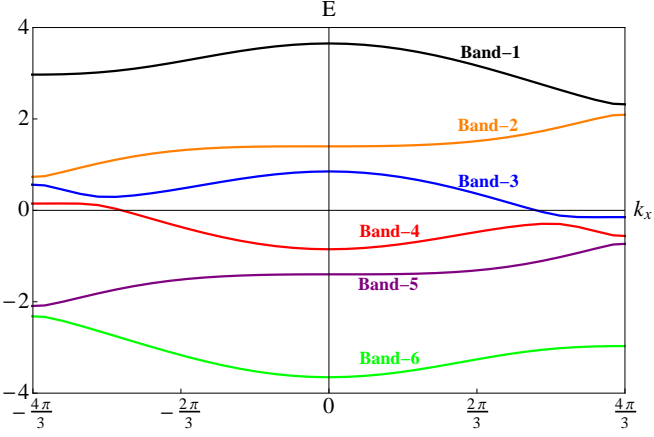}
\caption{ 
Illustration of the energy spectra of the $2m = 6$ bands along a cut with $k_y = 0$.  Here we take the same parameters as in Chua\etal,\cite{Chua10} $\{ J_\triangle, J_\nabla, J_\triangle^\prime, J_\nabla^\prime, J_5 \} = \{ 1.0, 0.3, 0.8, 0.5, 1.4 \}$.  Solving the $A_{IJ}$ matrix in Eq.~(\ref{Kitaev-type H}), there are $2m = 6$ bands which we label from top to bottom as 1 to 6.  Only half of the bands -- e.g., 1, 2, and 3 -- are used to solve the Majorana problem, as long as the others -- 4, 5, and 6 -- can be obtained via the relation between eigenvector-eigenenergy pairs such as $\{ \vec{v}_{4, {\bm k}}, \epsilon_{4, {\bm k}}\} = \{\vec{v}^*_{3, -{\bm k}}, -\epsilon_{3, -{\bm k}}\}$.
}
\label{bands spectrums}
\end{figure}

In the present case, $2m = 6$ and therefore three bands are sufficient to give us a full solution of the Majorana problem.  For an illustration of how all $2m = 6$ bands vary with momentum ${\bm k}$, we show the six bands in Fig.~\ref{bands spectrums} along a cut with $k_y=0$.  We label the bands from top to bottom as 1 to 6, and only bands 1 to 3 are used for the solution of the Majorana problem.  Specifically, we write the original Majoranas in terms of usual complex fermions as
\begin{eqnarray}
\nonumber && \Psi^M_I = \sum_{b=1}^3 \sum_{{\bm k}\in {\bf B.Z.}} \sqrt{2} \left[ V_{b,{\bm k}}(I) f_b({\bm k}) + V^*_{b,{\bm k}}(I) f^\dagger_b({\bm k}) \right]\\
&& = \sqrt{\frac{2}{N_{\rm uc}}} \sum_{b=1}^3 \sum_{{\bm k}\in {\bf B.Z.}} \left[ e^{i{\bm k} \cdot {\bm r}} v_{b,{\bm k}}(a) f_b({\bm k}) + \Hc \right] ~,\label{usual fermion}
\end{eqnarray}
where we used $V_{b,{\bm k}}(I\!=\!\{{\bm r},a\}) = \frac{1}{\sqrt{N_{\rm uc}}} v_{b,{\bm k}}(a) e^{i{\bm k} \cdot {\bm r}}$, ($N_{\rm uc}$ is the number of unit cells), and the complex fermion field $f$ satisfies the usual anti-commutaion relation, $\{ f^\dagger_b({\bm k}), f_b({\bm k}')\} = \delta_{bb'} \delta_{{\bm k}{\bm k}'}$.  In terms of the complex fermion fields, the Hamiltonian becomes
\begin{eqnarray}\label{usual_fermion_H}
\mathcal{H} = \sum_{b=1}^3 \sum_{{\bm k}\in {\bf B.Z.}} \epsilon_{b,{\bm k}} \left[ 2f^\dagger_b({\bm k}) f_b({\bm k}) - 1 \right] ~.
\end{eqnarray}

\begin{figure}[t]
  \includegraphics[width=\columnwidth]{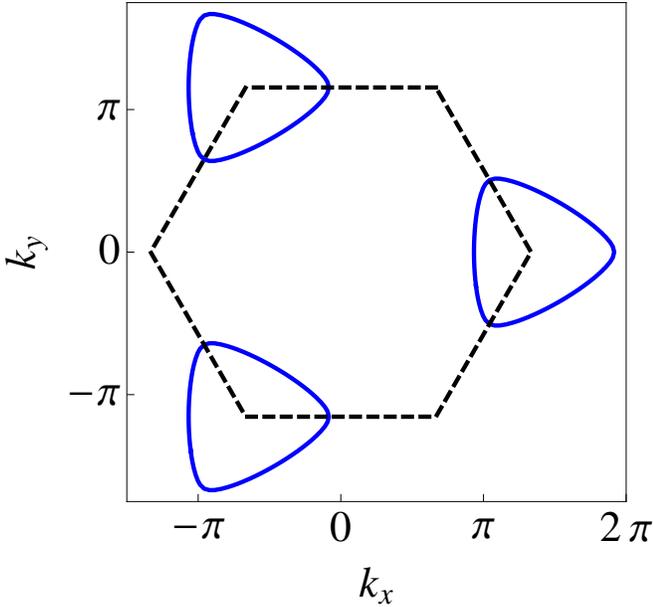}\\
  \caption{
Illustration of the Fermi pocket for true complex fermion $f_3$ in Eqs.~(\ref{usual fermion}),(\ref{usual_fermion_H}).  The parameters are the same as in Fig.~\ref{bands spectrums}  The hexagon is the Brillouin zone boundary.  Note that there is only one Fermi pocket for this set of parameters and the three pockets shown are related by a reciprocal lattice vector and are thus equivalent.}
\label{fermi surface}
\end{figure}

Considering the model parameters from Chua\etal\cite{Chua10} realizing the spin liquid with stable Fermi sea, we find that among these three bands, only $\epsilon_{3,{\bm k}}$ crosses the zero energy.  Hence, as far as the long-distance properties are concerned, we can retain only band 3 and its Fermi surface is shown in Fig.~\ref{fermi surface}.

\section{Bond Energy Correlators}\label{Sec:Bond correlator}
In this paper, we focus on the bond energy correlation functions.  There are several distinct bond energy operators one can consider.  However, all of them have similar long-distance behavior, so we present correlations for the bond energy operator corresponding to $J_\triangle$ term in the Hamiltonian between sites $1$ and $2$ as indicated in Fig.~\ref{Kagome Lattice} and defined as
\begin{eqnarray}
\mathcal{B}_{12c}({\bm r}) \equiv i J_\triangle u_{ij} c_i c_j
= i J_\triangle c_{{\bm r},1} c_{{\bm r},2}
= i J_\triangle \Psi^M_{{\bm r},1} \Psi^M_{{\bm r},2}, \label{def:bond energy}
\end{eqnarray}
where from the first to second equation we specified to the working gauge.  We will study bond energy correlator defined as
\begin{eqnarray}\label{def:bond-bond correlator}
G_B({\bm r}) \equiv \la \mathcal{B}_{12c}({\bm 0}) \mathcal{B}_{12c}({\bm r})\ra - \la B_{12c}({\bm 0}) \ra \la B_{12c}({\bm r}) \ra ~.
\end{eqnarray}
Power-law correlations in real space correspond to singularities in momentum space, which we can study by considering the structure factor
\begin{eqnarray}\label{structuref}
S_B({\bm q}) = \sum_{\bm r} G_B({\bm r}) e^{-i {\bm q} \cdot {\bm r}} ~.
\end{eqnarray}

We will present exact numerical calculation of the bond energy correlations in Sec.~\ref{Sec:Numerical data} using the definitions in Eqs.~(\ref{def:bond energy})-(\ref{structuref}).  Before showing the numerical data, we present a long wavelength analysis of such correlations due to the gapless Fermi sea of partons.

\subsection{Long wavelength analysis}\label{Sec:Long-wavelength analysis}
Focusing on the long distance behavior and therefore retaining only the contribution from band-3, the bond operator, Eq.~(\ref{def:bond energy}), can be written approximately as
\begin{eqnarray}\label{bond:general}
 \nonumber \mathcal{B}_{12c}({\bm r}) \simeq \sum_{{\bm k},{\bm k}'\in {\bf B.Z.}} \bigg{\{} \left[  M_{{\bm k}{\bm k}'} f_3({\bm k}) f_3({\bm k}') e^{i({\bm k} + {\bm k}') \cdot {\bm r}} + \Hc \right]\\
+ \left[ N_{{\bm k}{\bm k}'} f_3^\dagger({\bm k}) f_3({\bm k}') e^{-i({\bm k} - {\bm k}') \cdot {\bm r}} + \Hc \right] \bigg{\}},\hspace{0.6cm}
\end{eqnarray}
where $M_{{\bm k}{\bm k}'} = 2 i J_\triangle v_{3,{\bm k}}(1) v_{3,{\bm k}'}(2)/N_{\rm uc}$, $N_{{\bm k}{\bm k}'} = 2 i J_\triangle v^*_{3,{\bm k}}(1) v_{3,{\bm k}'}(2)/N_{\rm uc}$. 

In order to determine long-distance behavior at separation ${\bm r}$, we focus on patches near the Fermi surface of band 3 where the group velocity is parallel or antiparallel to the observation direction $\hat{\bm n} = {\bm r}/|{\bm r}|$, because at long distance $|{\bm r}| \gg k_F^{-1}$, the main contributions to the bond energy correlations come precisely from such patches.  Specifically, we introduce Right(R) and Left(L) Fermi patch fields and the corresponding energies
\begin{eqnarray}\label{fermi patches:momentum space}
&& f^{(\hat{\bm n})}_P(\delta {\bm k}) = f_3({\bm k}^{(\hat{\bm n})}_{FP} + \delta {\bm k}) ~,\\
&& \epsilon^{(\hat{\bm n})}_P (\delta {\bm k}) = |{\bm v}^{(\hat{\bm n})}_{FP}| \left( P \delta k_\parallel + \frac{\alpha^{(\hat{\bm n})}_P}{2} \delta k_\perp^2 \right) ~, \label{fermi patches:energy}
\end{eqnarray}
where the superscript $(\hat{\bm n})$ refers to the observation direction and $P = R/L = +/-$; ${\bm v}^{(\hat{\bm n})}_{FP}$ is the corresponding group velocity (parallel to $\hat{\bm n}$ for the Right patch and anti-parallel for the Left patch); $\alpha_{P = R/L}$ is the curvature of the Fermi surface at the Right/Left patch; $\delta k_\parallel$ and $\delta k_\perp$ are respectively components of $\delta {\bm k}$ parallel and perpendicular to $\hat{\bm n}$.  It is convenient to define slowly varying fields in real space
\begin{eqnarray}\label{fermi patches:real space}
f^{(\hat{\bm n})}_P({\bm r}) \sim \sum_{\delta {\bm k} \in {\rm Fermi~Patch}} f^{(\hat{\bm n})}_P(\delta {\bm k}) e^{i\delta{\bm k} \cdot {\bm r}} ~,
\end{eqnarray}
which vary slowly on the scale of the lattice spacing [and from now on we will drop the superscript $(\hat{\bm n})$].  Therefore, in this long wavelength analysis, the relevant terms in the bond operator are
\begin{eqnarray}
&& \hspace{-0.5cm} \mathcal{B}_{12c}({\bm r}) \sim \left[ \left(N_{RR} + N^*_{RR} \right) f_R^\dagger({\bm r}) f_R({\bm r}) + \left( R \rightarrow L\right) \right] ~\hspace{0.5cm} \label{q=0}\\
&& \hspace{-0.5cm} + \left[ \left(N_{LR} + N_{RL}^* \right) f^\dagger_L({\bm r}) f_R({\bm r}) e^{i({\bm k}_{FR} - {\bm k}_{FL}) \cdot {\bm r}} + \Hc \right] \label{q=R-L}\\
&& \hspace{-0.5cm} + \left[ \left(M_{RL} - M_{LR} \right) f_R({\bm r}) f_L({\bm r}) e^{i ({\bm k}_{FR} + {\bm k}_{FL}) \cdot {\bm r}} + \Hc \right], \label{q=R+L}
\end{eqnarray} 
where we dropped terms such as $f_R({\bm r}) f_R ({\bm r})$ due to Pauli exclusion principle.  The above long wavelength expression for the bond energy operator implies that the corresponding correlation function defined in Eq.~(\ref{def:bond-bond correlator}) contains contributions with ${\bm q} = {\bm 0}$, $\pm ({\bm k}_{FR} - {\bm k}_{FL})$, and $\pm ({\bm k}_{FR} + {\bm k}_{FL})$.  

More explicitly, for a patch specified by $\epsilon_P (\delta {\bm k})$ above, Eqs.~(\ref{fermi patches:momentum space})-(\ref{fermi patches:energy}), we can derive the Green's function for the continuum complex fermion fields as
\begin{eqnarray}\label{long-wavelength:green function}
\la f_{R/L}^\dagger({\bm 0}) f_{R/L}({\bm r}) \ra = \frac{\exp[\mp i \frac{3\pi}{4}]}{2^{3/2} \pi^{3/2} \alpha_{R/L}^{1/2} |{\bm r}|^{3/2}} ~.
\end{eqnarray}
Using this and the long-wavelength expression for bond energy operators, we can obtain the bond energy correlation, Eq.~(\ref{def:bond-bond correlator}),
\begin{eqnarray}
G_B({\bm r}) &\sim& -\frac{(N_{RR} + N^*_{RR})^2}{\alpha_R |{\bm r}|^3} - \frac{(N_{LL} + N^*_{LL})^2}{\alpha_L |{\bm r}|^3} \label{singular point:q=0}\\
&+& \frac{2 |N_{RL} + N^*_{LR}|^2 \sin[({\bm k}_{FR} - {\bm k}_{FL}) \cdot {\bm r}]}{\alpha_R^{1/2} \alpha_L^{1/2} |{\bm r}|^3} \label{singular line:R-L} \\
&+& \frac{2 |M_{RL} - M_{LR}|^2 \cos[({\bm k}_{FR} + {\bm k}_{FL}) \cdot {\bm r}]}{\alpha_R^{1/2} \alpha_L^{1/2} |{\bm r}|^3} . \label{singular line:R+L}
\end{eqnarray} 

Therefore, the above low energy description can be used to analyze the numerical data we obtain by exact calculations.  Here we also note that the model does not have inversion symmetry (and the time reversal is broken in the ground state), so the location of the corresponding R-L patches which are parallel or antiparallel to the observation direction can not be determined easily and need to be found numerically.

\subsection{Exact numerical calculation} \label{Sec:Numerical data}
\begin{figure}[t]
\includegraphics[width=\columnwidth]{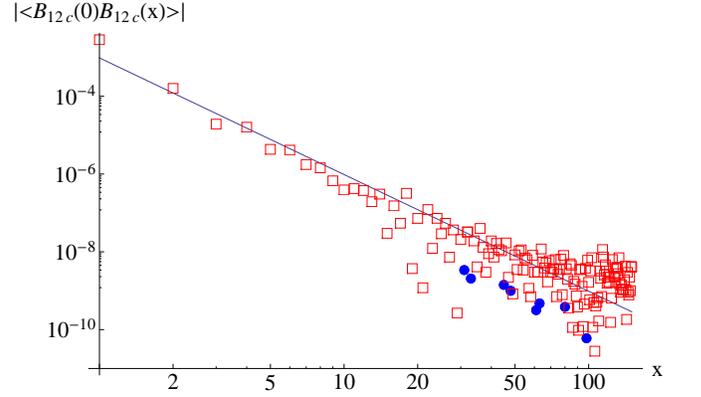}\\
\caption{
Illustration of power-law behavior of bond energy correlation function Eq.~(\ref{def:bond-bond correlator}).  We calculate $G_B({\bm r})$ with ${\bm r}$ taken along the $\hat{x}$-axis for a system containing $300 \times 300$ unit cells.  The log-log plot clearly shows $1/x^3$ envelope (straight line in the figure).  Here we show the absolute values $|G_B({\bm r})|$ and indicate the sign with open square boxes for negative correlations and filled circles for positive correlations. The irregular behavior is due to oscillating parts; these appear to be rather weak, but if we change the observation direction, the oscillating parts can be stronger.
}
\label{fig:bond-bond correlator}
\end{figure}

We calculate the bond energy correlations, Eq.~(\ref{def:bond-bond correlator}), for any real-space separation ${\bm r}$ and confirm that they have power law envelope $1/|{\bm r}|^3$.  For an illustration, we show the bond energy correlations for ${\bm r}$ along a specific direction, e.g.\ $\hat{x}$-axis, calculated on a $300 \times 300$ lattice.  In Fig.~\ref{fig:bond-bond correlator}, the log-log plot of $|G_B({\bm r})|$ along the $\hat{x}$-axis clearly shows the $1/|{\bm r}|^3$ envelope.  In addition, the irregular behavior of the data is due to oscillating components.  For certain directions, the oscillating parts are sufficiently strong that $G_B({\bm r})$ also changes signs.  The wavevectors of the real-space oscillations form some singular surfaces in the momentum space, which we will analyze next.

\begin{figure}[t]
\subfigure[Three-dimensional view of the structure factor]{\label{structuref:3dview} \includegraphics[width=\columnwidth]{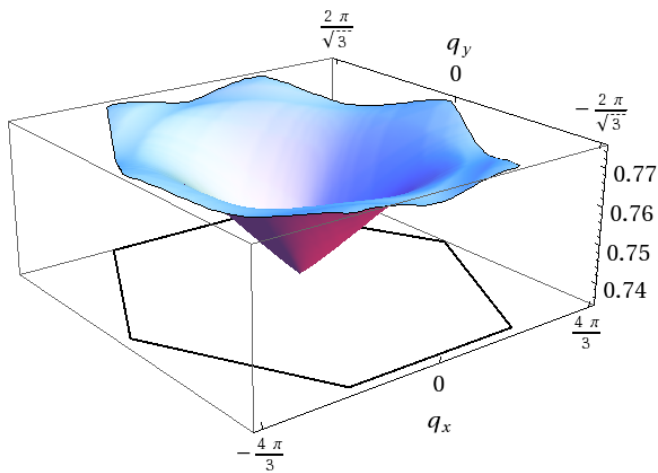}}
\subfigure[Top view of the structure factor]{\label{structuref:singular_surface}\includegraphics[width=\columnwidth]{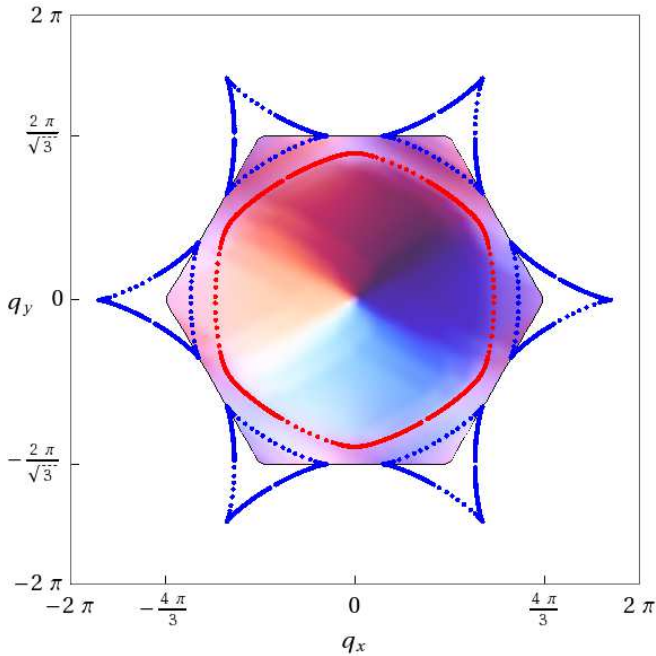}}
\caption{(a) 3D view of the structure factor of the bond energy correlation, $S_B({\bm q})$, defined in Eq.~(\ref{structuref}).  We can clearly see the singularity $S_B({\bm q}) \sim |{\bm q}|$ at ${\bm q} = 0$ and we also see weak singular lines, one forming a closed ring, and additional lines near the corners of the Brollouin zone.  
(b) These singular lines are brought out more clearly when the structure factor is viewed from top.  We also superposed the locations of the singularities calculated using the Fermi surface information:  The inner red ring specifies the line at ${\bm k}_{FR} - {\bm k}_{FL}$ and the outer blue triangles specify ${\bm k}_{FR} + {\bm k}_{FL}$ lines.}  
\label{structure factor}
\end{figure}

Shifting our focus on the structure factor $S_B({\bm q})$ defined in Eq.~(\ref{structuref}), we calculate the bond energy correlation at each site within a $100 \times 100$ lattice and numerically take Fourier transformation. Figure~\ref{structuref:3dview} gives a three-dimensional (3D) view of the structure factor.  We can clearly see cone-shaped singularity at ${\bm q} = {\bm 0}$, which is expected from Eq.~(\ref{singular point:q=0}), 
\begin{equation}
S_B({\bm q} \sim {\bm 0}) \sim |{\bm q}| ~.
\end{equation}

A closer look at the structure factor also reveals singular surfaces at ${\bm k}_{FR} - {\bm k}_{FL}$ and ${\bm k}_{FR} + {\bm k}_{FL}$, as expected from Eqs.~(\ref{singular line:R-L}) and (\ref{singular line:R+L}).
In order to see the location of the singular surfaces more clearly and compare it with our long wavelength analysis, we show top view of $S_B({\bm q})$ in Fig.~\ref{structuref:singular_surface}.  We also numerically calculate ${\bm Q}_{\pm} = {\bm k}_{FR} \pm {\bm k}_{FL}$ (by first finding corresponding Right and Left Fermi points with anti-parallel group velocities) and superpose these lines on the figure.  We can see that the lines we get from the long wavelength analysis match the singular features in the exact structure factor.  Note that the singularities are expected to be one-sided,
\begin{eqnarray}
S_B({\bm Q}_- + \delta {\bm q}) &\sim& |\delta q_\parallel |^{3/2} \Theta(- \delta q_{\parallel}) ~, \\
S_B({\bm Q}_+ + \delta {\bm q}) &\sim& |\delta q_\parallel |^{3/2} \Theta\left[- \delta q_\parallel \; {\rm sign}(\alpha_R - \alpha_L) \right].~~~~
\end{eqnarray}
The first line is singular from the inner side of the ``ring'' in Fig.~\ref{structuref:singular_surface} and the second line from the inner side of the ``triangles''.

\section{Conclusion}
We studied bond energy correlation functions in the Chua\etal\cite{Chua10} Kitaev-type model with a parton Fermi surface.  Unlike spin correlations, we found that the local energy correlations have power-law behavior in real space with an envelope of $1/|{\bm r}|^3$ and oscillations at incommensurate wavevectors that form singular surfaces in momentum space.  By combining low-energy theoretical analysis and exact numerical calculations, we determined the locations of the singular surfaces.  These bond energy correlations provide a gauge-invariant characterization of such gapless spin liquid.

We conclude by speculating about some interesting similarity with recent experiments in \dmit.\cite{Itou08, Itou10, MYamashita10}  While the thermal conductivity measurements\cite{MYamashita10} are consistent with the presence of a Fermi surface of fermionic excitations down to the lowest temperatures, very recent NMR experiments\cite{Itou10} show a drastic reduction in spin relaxation below temperature of the order $1$~K, almost as if a spin gap is opened.  This reminds of the present situation where the spin operators have short-range correlations, which occurs because some of the constituent partons have a gap (here are ultra-localized), while there remain partons that are metallic and give rise to metal-like thermodynamics and manifestly gapless properties such as the discussed local energy correlations.  Of course, the present model is on a different lattice and is very differently motivated.  However, in a recent paper\cite{SBMZeeman} working in a setting closer to the \dmit\ experiments, we discussed the following scenario in magnetic Zeeman field:  Upon writing the spin operator as $S^+ = f_\up^\dagger f_\dn$, we considered a state where one spinon species (say, $f_\up$) becomes gapped due to pairing, while the other species retains the Fermi surface.  In this case, $S^+$ spin correlations are short-range while the thermodynamics is metal-like.  Furthermore, just as in the present paper, there are other properties that are manifestly gapless, e.g., $S^z$ spin correlations and transverse spin-2 correlations.  It would be interesting to explore such scenarios in more realistic settings further.

\acknowledgments
This research is supported by the National Science Foundation through grant DMR-0907145 and by the A.~P.~Sloan Foundation.

\bibliography{biblio4KitaevFS}

\end{document}